# Sunspot characteristics at the onset of the Maunder Minimum based on the observations of Hevelius

V.M.S. Carrasco[1,2,3], J.M. Vaquero[2,4], M.C. Gallego[1,2], A. Muñoz-Jaramillo[3,5,6], G. de Toma[5], P. Galaviz[4], R. Arlt[7], V. Senthamizh Pavai[7], F. Sánchez-Bajo[8], J. Villalba Álvarez[9], J.M. Gómez[9]

[1] Departamento de Física, Universidad de Extremadura, 06071 Badajoz, Spain [e-mail: vmscarrasco@unex.es]

[2] Instituto Universitario de Investigación del Agua, Cambio Climático y Sostenibilidad (IACYS), Universidad de Extremadura, 06006 Badajoz, Spain

[3] Southwest Research Institute, Department of the Space Studies, Boulder, CO 80302, USA

[4] Departamento de Física, Universidad de Extremadura, 06800 Mérida, Spain

[5] High Altitude Observatory, National Center for Atmospheric Research, Boulder, CO 80301, USA

[6] National Solar Observatory, Boulder, CO 80303, USA

[7] Leibniz Institute for Astrophysics Potsdam, An der Sternwarte 16, 14482 Potsdam, Germany

[8] Departamento de Física Aplicada, Universidad de Extremadura, 06006 Badajoz, Spain

[9] Departamento de Ciencias de la Antigüedad, Universidad de Extremadura, 10003 Cáceres, Spain

**Abstract:** An analysis of the sunspot observations made by Hevelius during 1642–1645 is presented. These records are the only systematic sunspot observations just before the Maunder Minimum. We have studied different phenomena meticulously recorded by Hevelius after translating the original Latin texts. We re-evaluate the observations of sunspot groups by Hevelius during this period and obtain an average value 7% greater than that calculated from his observations given in the current group database. Furthermore, the average of the active day fraction obtained in this work from Hevelius' records previous to the Maunder Minimum is significantly greater than the solar activity level obtained from Hevelius' sunspot observations made during the Maunder Minimum (70% vs. 30%). We also present the butterfly diagram obtained from the sunspot positions recorded by Hevelius for the period 1642–1645. It can be seen that no hemispheric asymmetry exists during this interval, in contrast with the Maunder Minimum. Hevelius noted a ~3-month period that appeared to lack sunspots in early 1645 that gave the first hint of the impending Maunder Minimum. Recent studies claim that the Maunder Minimum was not a grand minimum period speculating that astronomers of that time, due to the Aristotelian ideas, did not record all sunspots that they observed, producing thus an underestimation of the solar activity level. However, we show the good quality of the sunspot records made by Hevelius indicates that his reports of sunspots were true to the observations.

## 1. Introduction

The systematic observations of sunspots started 400 years ago, in Galileo's time, when the telescope was used for the first time for astronomical purposes (Vaquero & Vázquez 2009, Muñoz-Jaramillo & Vaquero 2019). Since then, an outstanding episode of very low solar activity, commonly known as Maunder Minimum (MM), occurred during the period 1645–1715 approximately. This unusual period was mentioned first by Spoerer (1889) and Maunder (1890) in the end of the 19th century and the beginning of the 20th and confirmed by Eddy (1976) after reexamining the records available for that epoch.

Studies of solar activity carried out from cosmogenic isotopes have shown that periods as the MM (referred as grand minima) happened several times throughout the last ten millennia (Usoskin 2017). However, the MM is exceptional because it is the only grand minimum registered during the telescopic era. Thus, it has a great interest, for example, for the community of the solar physics and geophysics because it helps us to understand the influence of long-term solar activity on our planet and the entire heliosphere.

The modern understanding of the MM comes mainly through the effort made by Hoyt & Schatten (1998), who recovered thousands of sunspot observations carried out in that time. The database of Hoyt & Schatten (1998) has an observational coverage greater than 95% of the days considering the MM. However, Clette et al. (2014) showed that a considerable number of those records were extracted from solar meridian observations and thus they should not be combined with full disk observations. Vaquero & Gallego (2014) analyzed notes about sunspots in the astrometric observations made by the Royal Observatory of the Spanish Navy (San Fernando, Spain, 1833–1840) and concluded that this kind of sunspot records should be used with caution to reconstruct the past solar activity, For example, Vaquero & Gallego (2014) showed that many records were made with general and qualitative information and present a significant difference with respect to other sunspot observation made by Schwabe in the same days. Taken together, these studies suggest that the level of solar activity obtained from Hoyt & Schatten (1998) is likely underestimated. In fact, the last revisions of sunspot observations made by important astronomers who made observations during the MM have confirmed this idea (Carrasco et al. 2015; Carrasco & Vaquero 2016; Carrasco et al. 2019a). Taking this into account, Vaquero et al. (2016) created a revised collection of the number of sunspot groups including new information about sunspot records of the MM.

There is an ongoing debate about the true solar activity level during the MM. Hoyt & Schatten (1998) reported that sunspots were very rarely observed (around 2% of the days during entire period). However, modern works (e.g., Rek 2013; Zolotova & Ponyavin 2015) suggested that the MM was not a grand minimum of solar activity but a secular minimum. For example, Zolotova & Ponyavin (2015) argued that sunspots with irregular shapes were not recorded due to the ideas of that time about a perfect Sun. Nevertheless, Usoskin et al. (2015) and Gómez & Vaquero (2015) pointed out erroneous information in Zolotova & Ponyavin (2015) that led them to overestimate the solar activity level. Moreover, Carrasco et al. (2015), and Carrasco & Vaquero (2016), from the sunspot observations made by Hevelius and Flamsteed, obtained higher levels of solar activity than Hoyt & Schatten (1998) but compatible with a grand minimum of solar activity.

Several works have decisively contributed to our knowledge about the MM. For example, Spoerer (1889) was the first one who estimated heliographic latitudes of the sunspot recorded during the MM and noted the dominance of the southern hemisphere. Later, Ribes & Nesme-Ribes (1993) studied the solar cycle during the later phase of the MM and reconstructed the butterfly diagram from heliographic latitude records showing a strong hemispheric asymmetry since sunspots were mainly observed in the southern hemisphere. Vaquero et al. (2015a) used the data base of Hoyt & Schatten (1998) to obtain a cycle length of 9 ± 1 years for the core of the MM (1650–1700) and an average sunspot number, based on the active day fraction method, to be below 5-10 for this 50-year interval *versus* 0.35 determined from Hoyt & Schatten (1998) for that same period. Furthermore, Vaquero et al. (2011) recovered information not analyzed previously about sunspot records made before the MM and determined a lower solar activity level than previously estimated for the solar cycle preceding the MM reducing cycle peak sunspot numbers of ~70 for 1638 and 1639 to ~20. This result implies that the transition from normal to low solar activity level was not abrupt, but rather was similar to the transition at the end of the MM when the low activity regime was gradually changing to normal solar activity regime.

The aim of this work is to analyze the sunspot records made by Johannes Hevelius just before the MM. In Section 2, we present the sunspot observations recorded by Hevelius in *Selenographia*. Section 3 describes the observational methodology and, insofar as possible, instrumentation used by Hevelius. In Section 4, we explain some particular phenomena observed by Hevelius. Section 5 shows the estimation of the solar activity derived from these records. Section 6 deals with comments on the phenomenology and understanding of sunspots and in Section 7, the main conclusions are given.

## 2. Sunspot observations by Hevelius: metadata

*Selenographia* is a famous astronomical work, published in 1647, that was written by the astronomer Johannes Hevelius (1611–1687). It includes important astronomical observations mainly related with the cartography of the Moon. In addition, sunspot observations carried out by Hevelius at Danzig (currently Gdansk, Poland) during the period 1642-1645 can be found in the appendix of this work. These sunspot observations are of special interest for two main reasons. First, the observations are of outstanding quality, made in a professional observatory including explanatory text and drawings. Secondly, they are the only available regular and continuous sunspot records at the beginning of the MM available to us. For this particular period (1642-1645), in addition to the 14 observation days recorded by Rheita in 1642 (Gomez & Vaquero 2015), only two additional possible daily sunspot observations are available (Hoyt & Schatten 1998), one observation day recorded by Linemanns (in 1644) and the another one by Gassendi (in 1645). For comparison, Hevelius (1647) reported observations for sunspots on 17 days in 1642, 135 in 1643, 168 in 1644, and 4 in 1645. Although *Selenographia* was previously consulted by Hoyt & Schatten (1998) for the construction of the Group Sunspot Number index, we present here an analysis of the sunspot observations recorded in this historical source after translating all the annotations made by Hevelius from the original Latin text into English language.

In an appendix of *Selenographia*, a diary of sunspot observations is provided by Hevelius, including drawings of the solar disc that contain the apparent trajectories of sunspots crossing the solar disc (Figure 1). As it can be seen in Figure 1, Hevelius includes in the drawings the following details: i) the line crossing the center of the drawing marks the ecliptic; ii) the eastern and western solar limb are pointed out as *A* and *B*, respectively; iii) a different letter is assigned to each sunspot group; iv) the numbers close to sunspots indicate the observation date; and v) tables list information about the month (entire name or abbreviation of the month, in Latin), day (*D*), hour (*Hor*, indicating if the observations were carried out before (*m*) or after (*u*) noon), and minute (′) of the observations as well as the angle of the ecliptic (*Ang*). In addition to the drawings, Hevelius provided daily detailed comments with the description of each observation. We want to note that, in general, the information included both the drawings and notes coincide. However, sometimes it is necessary to consult both the diary and drawings to have a complete information about sunspots observed each day. Specifically, information about spotless days is only included in the diary, there are descriptions of sunspots in the diary which are not represented in the drawings (such as on 24 June 1643), and there are sunspots represented in the drawings but not described in the diary of observations (for example, group “e” in the drawing V on 25 June 1643).

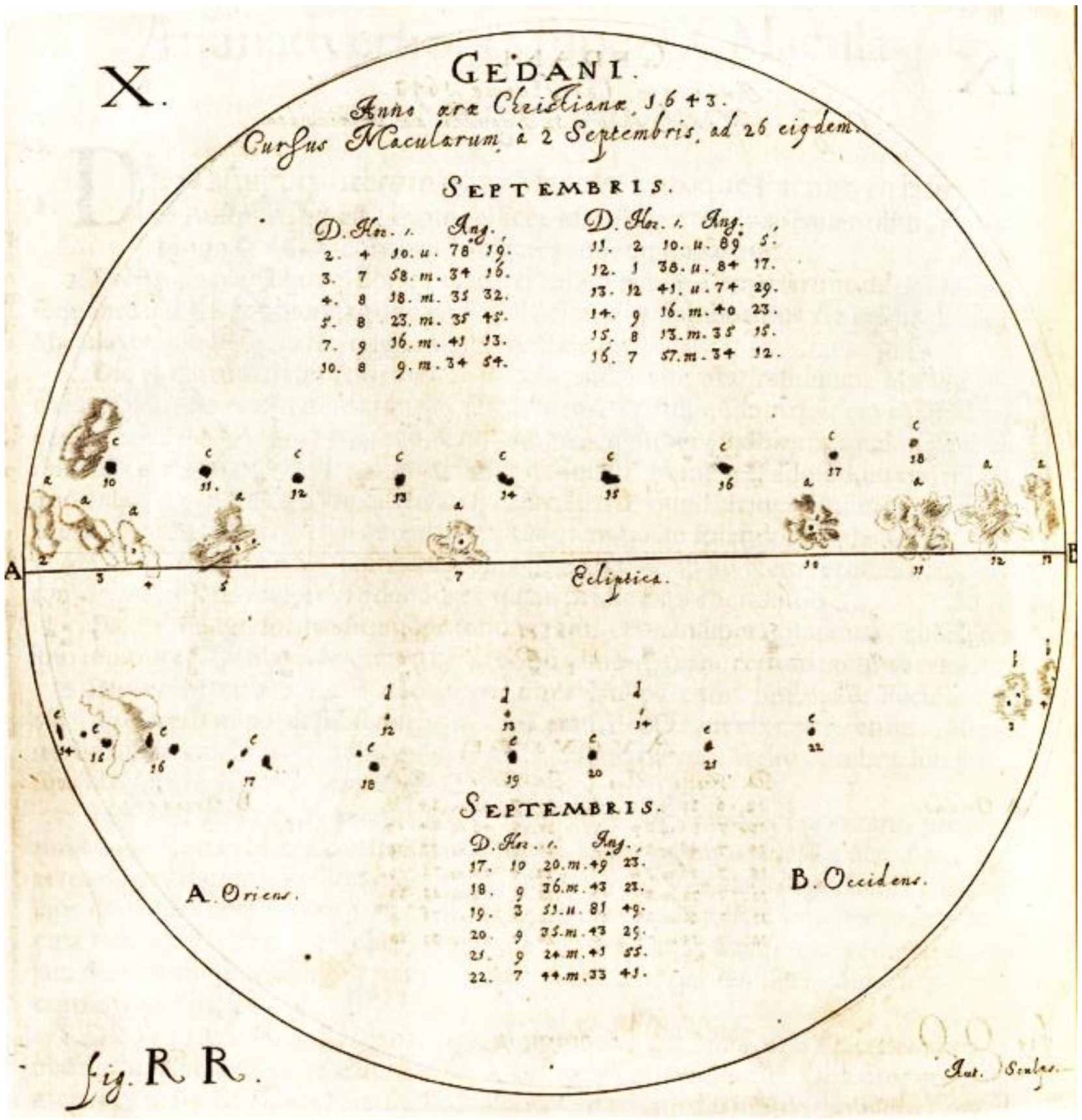

Figure 1. An example page of *Selenographia* with a sunspot drawing made by Hevelius on September 1643 [Source: *Selenographia*, courtesy of the Library of the Astronomical Observatory of the Spanish Navy].

In his daily records, Hevelius provided a detailed description about the sunspots observed by him. Information as the sunspot morphology, size, color, or movement is included in these descriptions. In addition to sunspots, Hevelius also observed and recorded other phenomena as faculae, umbrae, and "halos". Details about these phenomena are shown in next sections. Furthermore, Hevelius provided information about the trajectory of different sunspots crossing the solar disc. Hevelius checked and noted that, for example, the trajectories of sunspots were rectilinear in the solstices and curved around the equinoxes. In this work, we do not pay attention to the study of the trajectory of sunspots recorded by Hevelius but these descriptions could be of interest, for example, to the historians of the science for studies about heliocentrism (Smith 1985; Mueller 2000). For this reason, we highlight the following annotation made on 14 April 1644 where it can be seen that Hevelius (1647, p. 516) realized sunspots follow the same trajectories during the same months: original text – "[…] Ex quibus autem omnibus Macularum motus plane esse regularem atque omni tempore constantissimum, nimis quam aperte innotescit. Quamvis vero Maculae hae neutiquam perennent, motus tamen eius est uniformis, sic ut in iisdem similibus Mensibus, semper eadem motuum curvitas […]"; English translation – "[…] From all this, it is clear that the movement of the spots is regular and always very constant. Although sunspots are not permanent, their movement is uniform, so, in the same months, the curve describing their movements is always the same […]".

## 3. Observation method

Hevelius (1647, p. 99) described different methods to observe sunspots. For example, he explained that the most noticeable sunspots can be observed by naked eye and camera obscura. Hevelius even indicated that sunspots can be appreciated without causing damage to the eye if among two colored crystals, a lighter color sheet pierced with a fine needle is interposed and all glued with wax or bitumen (very well-known in the history of astronomy to observe solar eclipses). According to Hevelius, the best and most comfortable method to observe sunspots is by telescope and the most effective method is using a helioscopic machine, invented by Scheiner and improved by Hevelius to allow one person to observe sunspots without help from another (Kampa 2018). Hevelius provided a detailed description of the different parts of the helioscopic machine and the procedure used to use this instrument (Figure 2). Hevelius pointed out that observations should be carried out in a dark room with windows of wood where one of these has a round hole to incorporate a freely movable brass or wood sphere of about 0.15-0.17 meter in diameter (i.e., 8 or 9 digits where 1 digit is equal to 19 mm). In the center of this sphere a hole was made to incorporate a 0.45 m brass tube where telescopes were introduced to perform the observations. We did not find the information about the focal length of the particular helioscope used by Hevelius to observe sunspots but Kampa (2018) indicated that it was about 0.6 m. Thus, the large telescopes for which Hevelius became famous, were not used for solar observations. The diameter of the lens used by Hevelius remains unknown. However, it should be around 0.05 m since that is the diameter of the brass tube where telescopes were placed. Another extensible tube around 2.5-2.75 m in length connected the sphere with a wood bench where the sunspot drawing was made on a white sheet glued with wax to a table. The bench also contained vertical threaded rods to move the table with the observation sheet up or down. The observation sheet must be placed orthogonal to the solar rays. When the projected image of the Sun was greater or smaller than the solar disc drawn in the observation sheet, Hevelius indicated that either the table or the telescope were adjusted until the projected image of the Sun fitted the outline of the solar disc. Finally, Hevelius noted that to perform a perfect sunspot record required a sundial divided, at least, every three minutes where the meridian line is well marked (see left center part of Figure 2). He also indicated that sunspots must be marked with a lead stamp indicating the day and time when the observations were made. The quality of the telescope and the method used by Hevelius were appreciated, for example, by Gassendi (Olhoff 1683) and Linemann (1654).

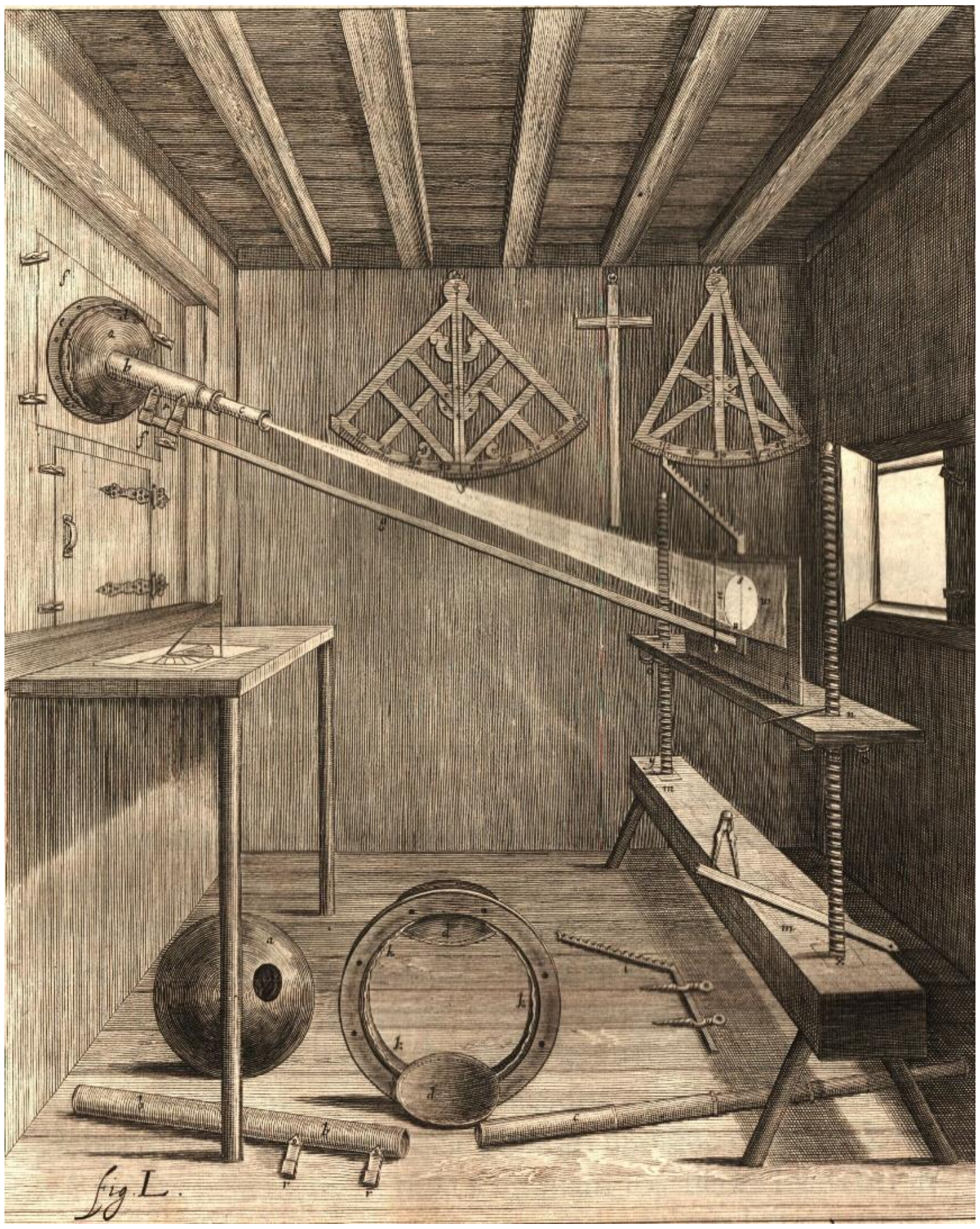

Figure 2. An image of the Hevelius' observatory [Source: *Selenographia* (Hevelius 1647), courtesy of the Library of the Astronomical Observatory of the Spanish Navy].

Hevelius (1647, p. 102) pointed out what a correct drawing of the sunspots required. He stated that one observer should not be satisfied with a single daily observation, but he should check again over the course of the day if there were new spots, as well as if their sizes, appearances and shapes had changed, or if they were more compact or more diffused, and if they had a nucleus or not. Hevelius also indicated that faculae must be noted, if they appear. Moreover, He says: Original text: "[…] quemadmodum frequens usus hoc quemlibet docebit et idem ex meis observationibus sit conspicuum… Postquam igitur uno die, maculae Solis cum genuinis coloribus, umbris et faculis recte sunt notatae, tum de die in diem, sudo existente coelo, ille labor est continuandus ut ex hisce accuratis observationibus macularum, cursus et mirabilis mutatio deprehendi possit."; English translation: "[…] The frequent practice will teach anyone how this is done and it can also be seen well from my observations… So, after having recorded sunspots in one day with their natural colors, with shadows and faculae, then every day that sky is clear the same task will continue so that, from these rigorous observations, its path and its admirable change can be seen" (Hevelius 1647, p. 102-103). Furthermore, Hevelius clarified that the image of the Sun is inverted on the observation sheet and therefore it must be turned upside down. Hevelius also noted that the sunspot positions in the sunspot drawings published in *Selenographia* are corrected for this effect.

Hevelius (1647, p. 103) indicated that his observation method was better than that used by Scheiner because, according to Hevelius, his machine was easier to use and therefore he could hold a stable image of solar disk on the observation sheet without difficulty. Moreover, Hevelius built a table customized for his observation site (Gdansk, Poland) to establish the location of the ecliptic at every moment of the year on the observation sheet. Finally, Hevelius indicated that sunspots should project on a single observing sheet while they move across the Sun and that "Certainly, this method of recording the motion of the sunspots on the Sun is labor-intensive and tedious, but the effort and enthusiasm to properly show the movement of these should help to tolerate and overcome any annoyance" [Original text: "[…] Hic quidem modus recte exprimendi curriculum macularum subter Solem cum labore et taedio est coniunctus: conatus et alacritas tamen legitime exhibendi motum earum omnem molestiam debet tolerare et superare”] (Hevelius 1647, p. 105).

## 4. Observational phenomena recorded by Hevelius

### 4.1. Bright halos surrounding sunspots

In the descriptions of his sunspot observations, Hevelius provided comprehensive details about several phenomena that he observed on the Sun although they were not well-known in that epoch. An intriguing phenomenon observed by Hevelius was sunspots surrounded by luminous “halos”. Hevelius presented this kind of phenomenon in seven of his sunspot drawings: i) October-November 1642, ii) November 1643, iii) April 1644, iv) and v) May 1644 (2 cases), vi) June 1644, and vii) July 1644. Maybe, the most striking case was recorded on May 1644 (Figure 3A) when Hevelius stated: original text – “[…] d imprimis enim non solum multo maior haloneque splendidissimo coronata deprehensa, sed et nucleum in eius meditullio densissimum nigerrimumque nec non duas maculas, ex superiori umbra recenter procreatas, exhibuit. Qua certe ampliorem atque magis egregiam Macula videlicet d, vix memini me, longo elapso temporis spatio, observasse.”; English translation – “[…] the sunspot “d” was not only seen much larger and surrounded by a luminous halo, but it also showed a very dense black nucleus in its core, as well as two spots formed shortly before from an earlier umbra. Certainly, I cannot remember seeing in a long time a bigger and more striking sunspot than this sunspot “d””. It can be seen in the previous description that Hevelius recorded three different tonalities in the observed sunspot “d”: a dark nucleus inside the sunspot and a halo around it. We highlight that the observation of a halo surrounding a sunspot is a singular phenomenon, especially in the first telescopic sunspot observations. These bright halos were recorded by Hevelius around medium-large and isolated sunspots. Some works have shown the existence of faint rings surroundings several large isolated sunspots (Rast et al. 1999, 2001) that are about 0.5-1.0% brighter than the surrounding photosphere and have a thermal origin. However, the existence of these non-magnetic bright rings has been confirmed only in carefully calibrated images from the Precision Solar Photometric Telescope (Rast et al. 1999, 2001).

Alternatively, the halos recorded by Hevelius surrounding some sunspots could represent faculae, despite the fact that Hevelius perfectly knew the concept of faculae and refers to faculae often in *Selenographia*. The sunspot named AR12546 by NOAA (National Oceanic and Atmospheric Administration) is shown in Figure 3B and 3C from modern images taken by the *Helioseismic and Magnetic Imager* (HMI) on board the *Solar Dynamics Observatory* on 24 May 2016 in the visible continuum. On 22 May 2016 (Figure 3B) (Carrasco et al. 2018), it was located at 7º S and 35º W, a longitude comparable to the one of sunspot "d" in the Hevelius drawing of Figure 3A. We note that the facula around the sunspot is not visible at this disk position. However, it was visible two days later. Figure 3C shows the observation of the same sunspot two days later, near the limb with umbra and penumbra. Its coordinates respects to the central meridian of the solar disk were 7º S and 63º W. The nearly circular facula around the sunspot is reminiscent of the halo drawn by Hevelius. However, faculae are visible in the HMI continuum images only when they are close to the limb. Therefore, it seems unlikely those halos are faculae because Hevelius draws the halos as separate features from the faculae for sunspots near the limb, as shown in Figure 4 (left panel), which indicates he considered the halo a different phenomenon. Furthermore, we know that faculae are clearly visible only near the limb while in *Selenographia* there are sunspots that were very close to or at disk center that were drawn and described by Hevelius with a halo around them, like the sunspots observed on May 22 and 23, 1644 (Figure 4, left panel) and on May 10 and 11, 1644 (Figure 4, right panel). Another interpretation could be that the nucleus represented the darkest part of the umbra and the halo the lighter penumbra but this would make the already large sunspot drawn in Figures 5 exceptionally. Thus, regardless of whether Hevelius halos represent a magnetic or non-magnetic brightening, it remains unclear to us what would be their modern counterpart and how Hevelius would have been able to detect these halos so clearly with his helioscope. In closing, we remark that sunspots "b" (Figure 4, left panel) and "d" (Figure 4, right panel) are very large spots, around 1300 and 500 millionths of solar hemisphere respectively, that were observed only months before the onset of the MM.

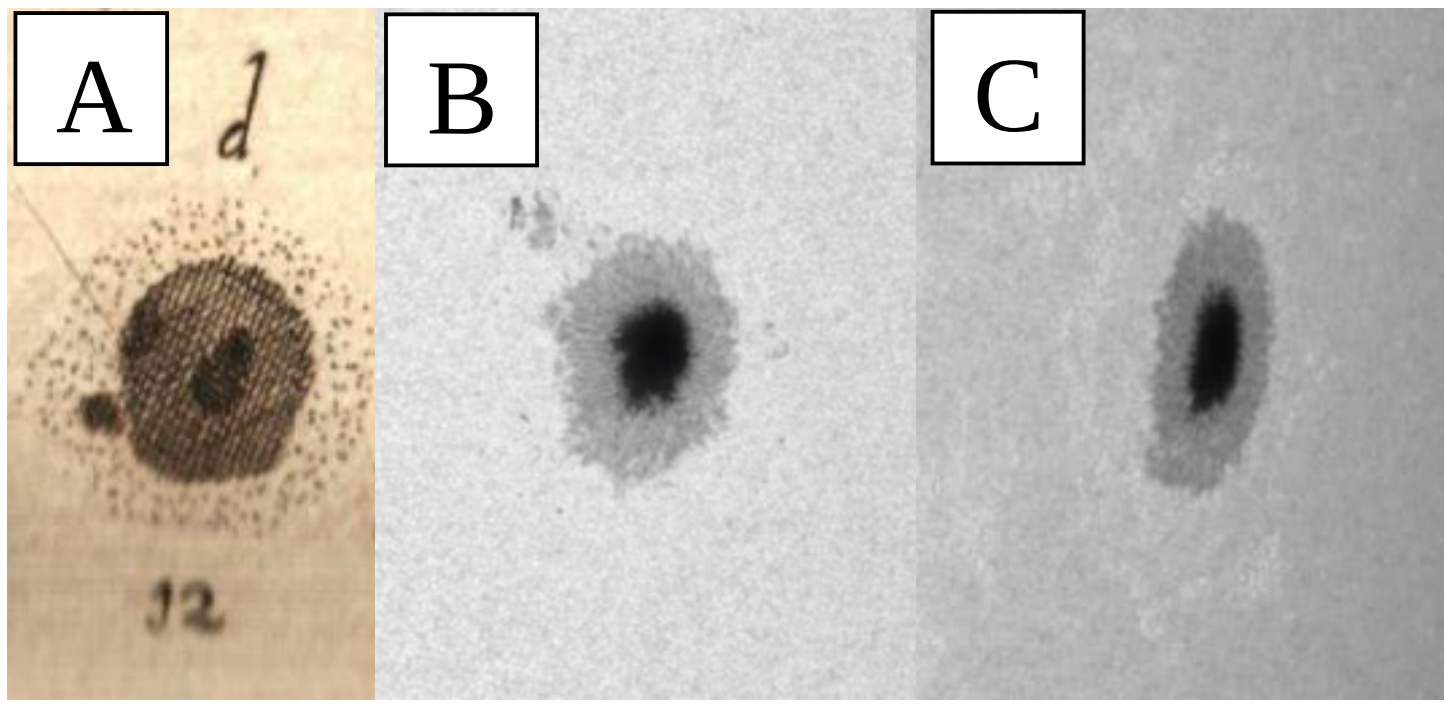

Figure 3. Sunspot observed by (a) Hevelius surrounded by a luminous halo on 12 May 1644 and [Source: *Selenographia* (Hevelius 1647), courtesy of the Library of the Astronomical Observatory of the Spanish Navy] and (b-c) modern sunspot images taken by the HMI instrument on board the *Solar Dynamics Observatory* in the visible continuum for the same sunspot on 22 May 2016 at 23:28UT and 24 May 2016 at 23:28UT, respectively [Source: http://suntoday.lmsal.com/]. The sunspot in the center panel is located at a longitude comparable to the sunspots in the Hevelius drawing shown in the left panel. The sunspot in the right panel is closer to the limb.

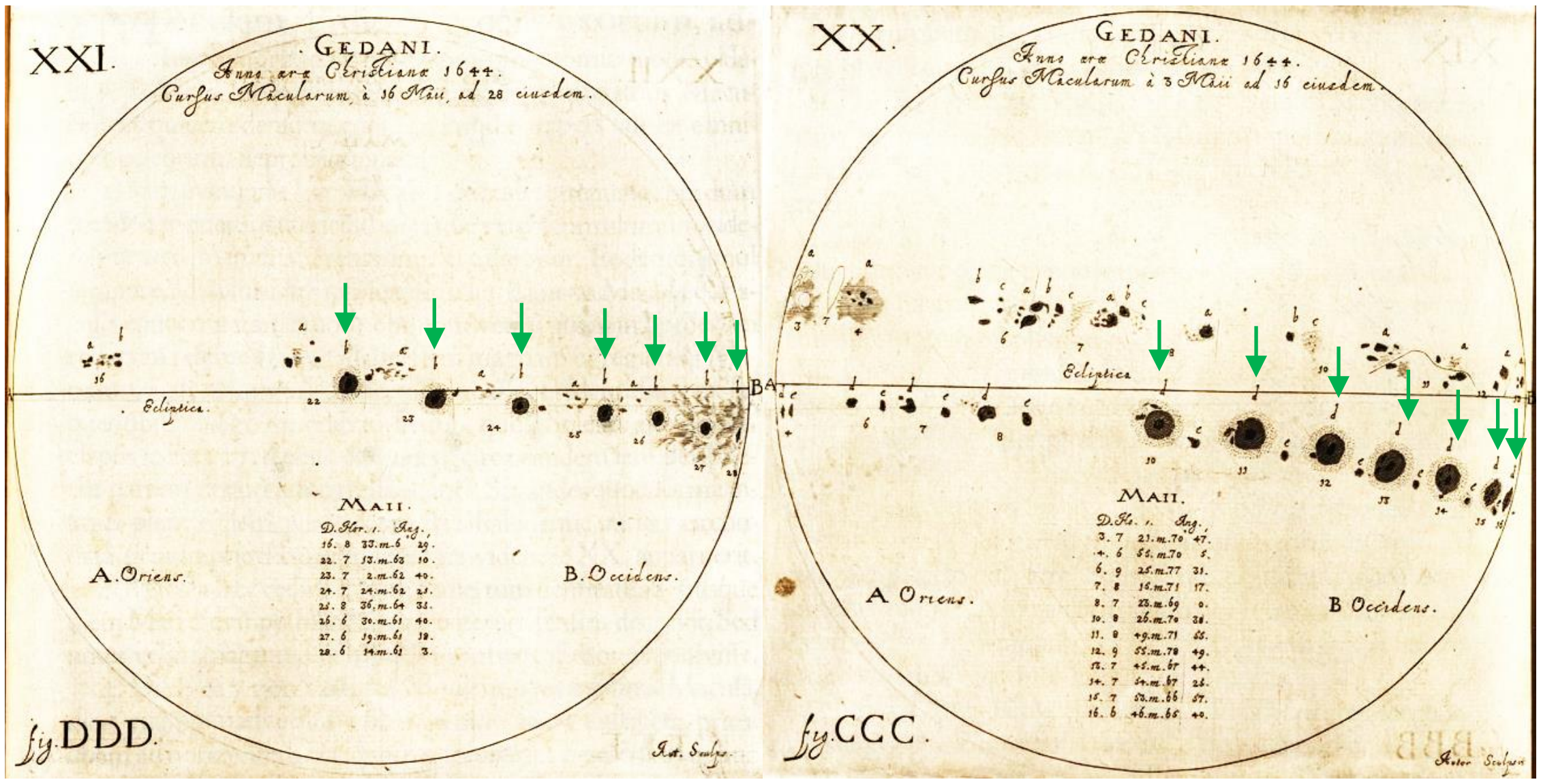

Figure 4. (Left panel) Example of a sunspot (see sunspot "b" with green arrows) observed by Hevelius surrounded by a halo on 25-28 May 1644 [Source: *Selenographia* (Hevelius 1647)]. The halo is clearly depicted as a separate feature from the facula and the facula is drawn only near the limb. (Right panel) Observing sheet showing the evolution of sunspot "d" (see green arrows) as it crosses the solar disk on May 10-16, 1644. Note that the halo

drawn around the large sunspot is present at all disk positions and there are not significant differences in the size and shape of the halo as the sunspot moves from disk center toward the solar limb [Source: *Selenographia* (Hevelius 1647)].

### 4.2. Umbrae and faculae

A term very frequently used by Hevelius in his annotations was "umbra". In most cases, Hevelius employed "umbra" to assign the darker areas in facular regions. On the one hand, we highlight a noticeable umbra observed by Hevelius on 25 September 1643 (Figure 4A). On the other hand, it is worthy of mention that Hevelius recorded cases where great areas of the solar photosphere were covered by "faculae and umbrae". The most remarkable case can be found on 20 July 1643 (Figure 4B): original text – "[…] Quae quidem sex Maculae, hac ipsa 20 scilicet die, comitatum mirificum Facularum umbrarumque post se trahebant, dum ille tertiam Solaris diametri partem longitudine, latitudine autem nonam dictae diametri partem aequabat […]"; English translation – "[…] These six sunspots have an amazing accompaniment of faculae and umbrae on 20th. This accompaniment was equal to a third of the solar diameter in length and a ninth in latitude. And that is something that certainly deserves to be highlighted […]". Moreover, Hevelius also used this term to define dark regions placed between sunspots (Figure 4C). This case is more confusing but it could refer to darker regions between sunspots due to a contrast effect between the faculae surrounding sunspots and photosphere.

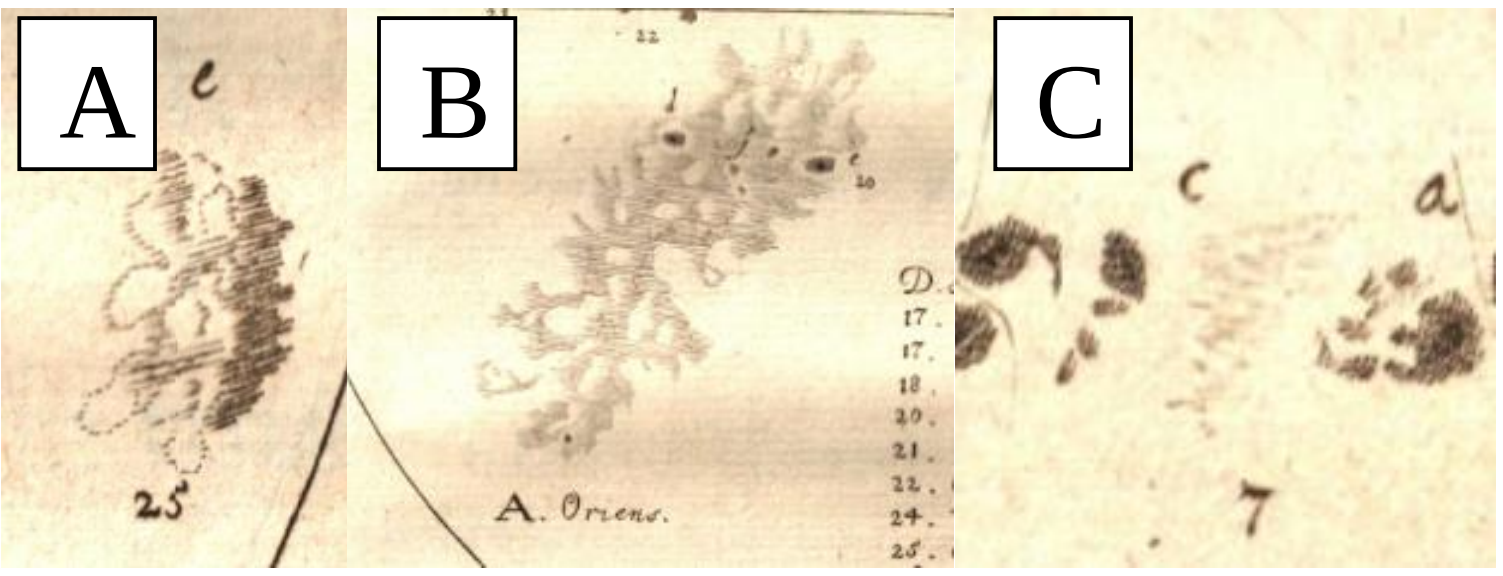

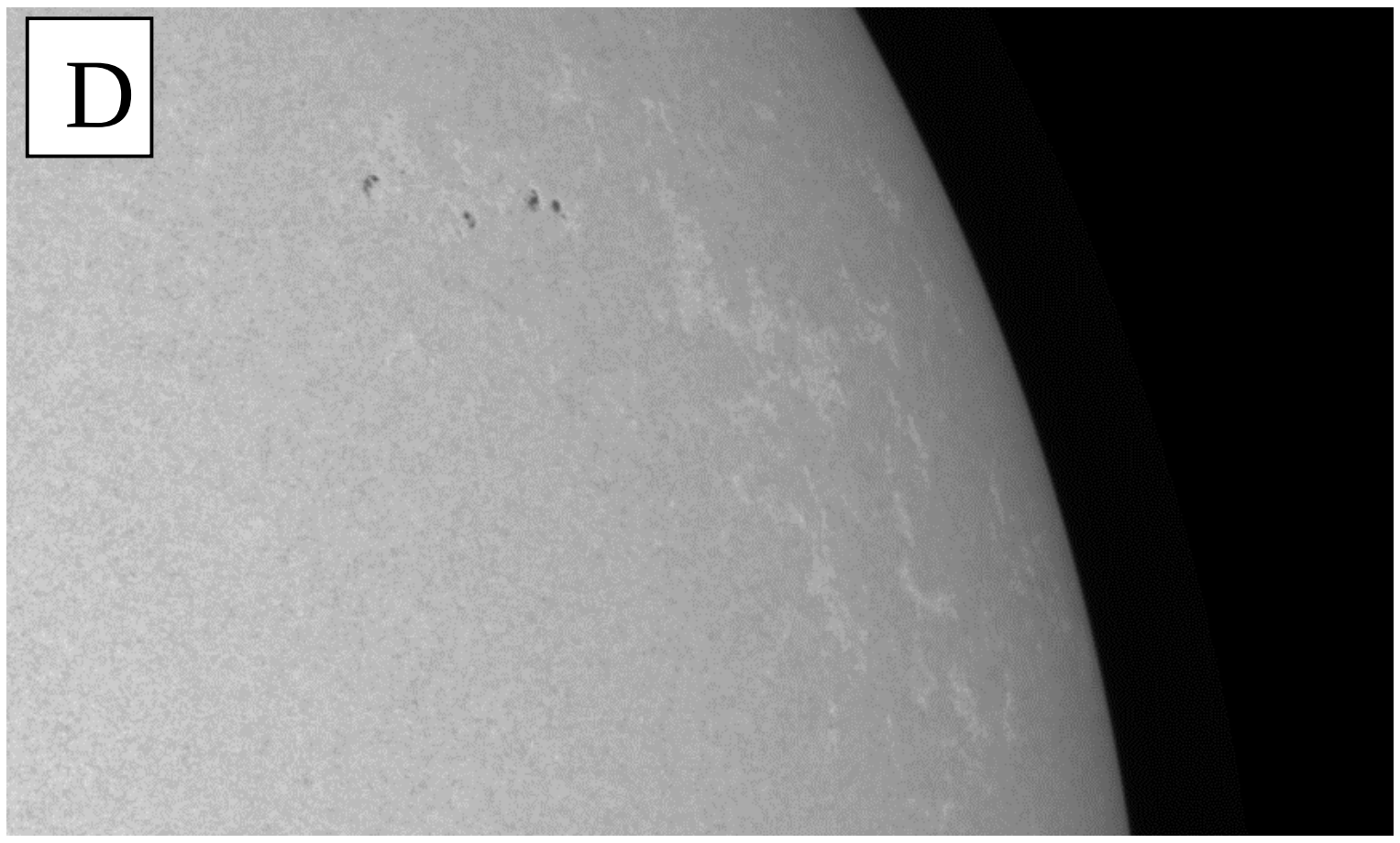

Figure 4. (a) A noticeable "umbra" (25 September 1643), (b) great accompaniment of faculae and umbrae (20 July 1643), and (c) "umbra" placed between sunspots (7 May 1644) recorded by Hevelius [Source: *Selenographia* (Hevelius 1647), courtesy of the Library of the Astronomical Observatory of the Spanish Navy], and (d) a modern example of a great accompaniment of faculae in images taken by the HMI instrument on board the *Solar Dynamics Observatory* on 7 March 2014 at 23:58:13.700 [Source: http://suntoday.lmsal.com].

4.3. Fogs and clouds

Hevelius sporadically recorded in his annotations the observation of "foggy" (*nebulisque*) spots or "clouds" (*nimbo*) around sunspots. For example, in the second sunspot drawing recorded in *Selenographia* (6 November 1642), Hevelius (1647, p. 501) used the term "fog" (nebula) to describe what is clearly a facula region (Figure 5A): original text – "Unica tantum Macula c, umbris nebulisque coronata spectabatur […]"; English translation – "One only "c" ["b" in the drawing] sunspot was observed surrounded of umbrae and fogs […]". However, when that same sunspot was in the opposite solar limb (Figure 5B), Hevelius did not use the term "umbrae and fogs" but "umbrae and faculae" to describe the region located around the sunspot. Hevelius (1647, p. 504) also employed the term "fog" (*nebule*) to refer probably to penumbrae zone of a sunspot observed on 26 June 1643 (Figure 5C): original text – "Et una quidem illarum, quae alteri adhaerebat, satis quidem magna, sed admodum rara ac sparsa, instar debilis nebulae diluta comparuit, ita ut passim Solem splendentem per illam clare deprehenderem […]"; English translation

– “One of them [sunspots], quite large but very tenuous and scattered, which was attached to another sunspot, was diluted as a weak fog, until, through it, it could be clearly seen the Sun shine in all its splendor […]”. On 23 May 1643, Hevelius (1647, p. 502) indicated that the sunspot “a” seemed surrounded by a cloud (*nimbus*) (Figure 5D). In this last case, this “cloud” is probably a facula next to the sunspot. Moreover, Hevelius (1647, p. 507) used the term “cloud” (*nimbus*) on 3 August 1643 to describe what seems the penumbra of that sunspot (Figure 5E): original text – “Adhuc alia valde parvula, debilis recensque nata Maculam a sequebatur, quae cum maiore, nimbo coronabatur, ut et die subsequente […]”; English translation – “Another very tiny, weak, and newly created sunspot accompanied the sunspot “a”; it was surrounded by a cloud, along with the larger one; and it happened the same the next day […]”. It is difficult to identify that phenomon with modern examples. Figures 5F, 5G, and 5H show modern examples corresponding to the phenomena shown in Figure 5A-B, C, and D according to the terms described by Hevelius as “fogs” and “clouds”.

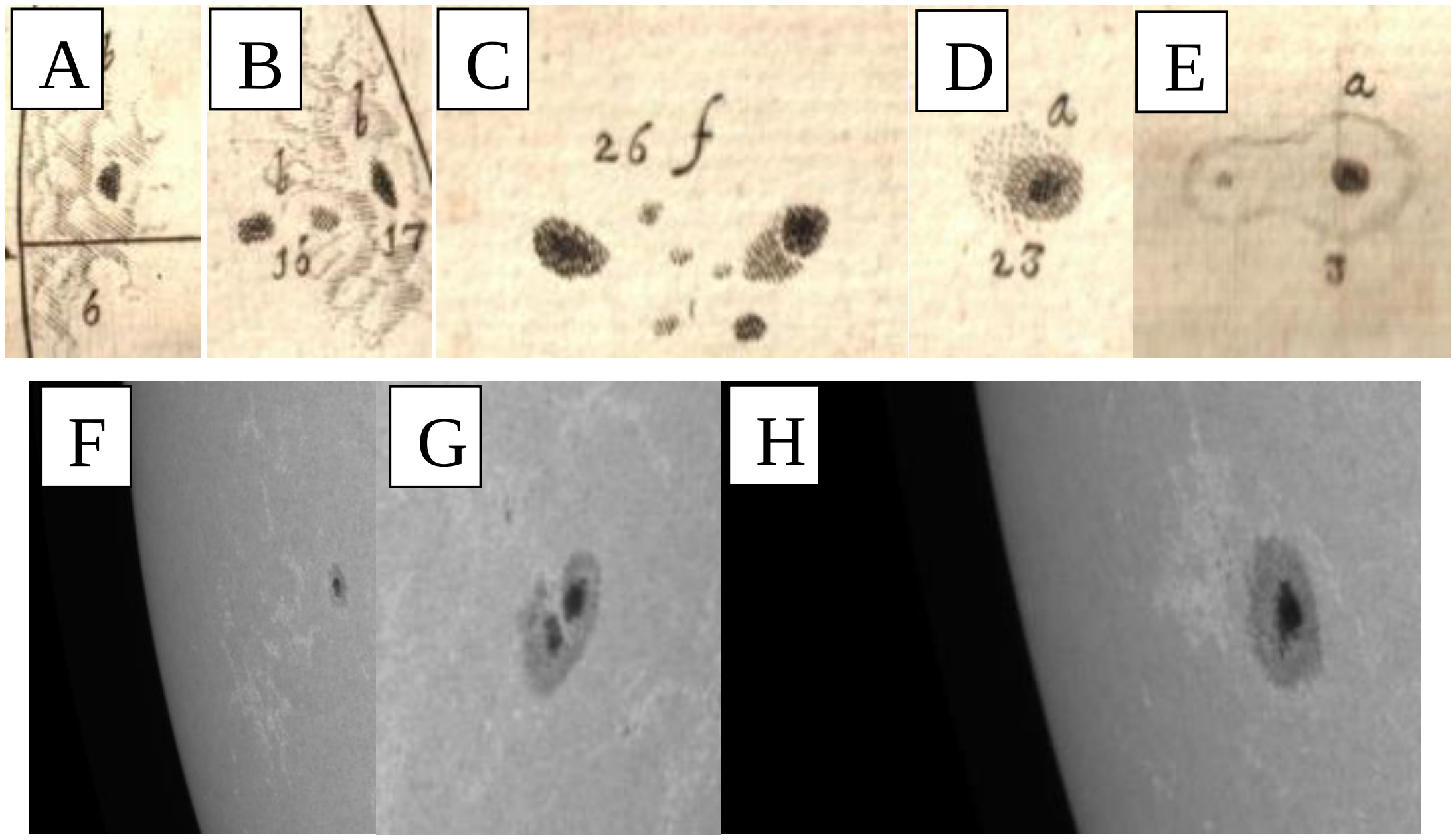

Figure 5. Sunspots accompanied by “fog” (*nebulisque and nebula*) (A- 6 November 1642, B- 17 November 1642, and C- 26 June 1643) and “cloud” (*nimbus*) (D- 23 May 1643, and E- 3 August 1643) according to Hevelius’ descriptions [Source: *Selenographia* (Hevelius 1647), courtesy of the Library of the Astronomical Observatory of the Spanish Navy]. Panels (F), (G), and (H) represent modern examples of possible “fogs” and “clouds” in continuum images taken by the HMI instrument on board the *Solar Dynamics Observatory* on 4 May 2014 at 23:58:21.700, 18 May 2014 at 23:58:23.300, and 25 August 2013 at 23:58:22.700, respectively [Source: http://suntoday.lmsal.com].

4.4. Formation and rotation of sunspots

Hevelius (1647, p. 501) observed that, on the one hand, sunspots can appear and disappear at the solar limb and, on the other hand, sunspots are also born and die on the visible hemisphere of the Sun (9 November 1642): original text – “[…] Unde equidem satis elucet Maculas hasce Solem non fuisse subingressas, sed in medio eius disco natas dissipatasque […]”; English translation – “[…] From which it can be deduced without doubt that these sunspots did not timidly enter the Sun, but they were born and dissipated in the middle of the solar disc […]”. This fact led Hevelius to ask why some sunspots remained for more time on the solar disc than other ones. For example, on 23 August 1643, Hevelius (1647, p. 508) registered: original text – “[…] Ex pridianis recens natis, tantummodo adhuc una f erat superstes, altera g omnino iterum exstincta, sic ut nec tenue sui indicium reliquerit. Quaeritur ergo hic merito. Cur haec dissipata Macula, ut et altera f, non tam diu in Sole duraverit, quam illa c, cum tamen aeque magna densa ac spissa qua materiam pridie sit observata. c tamen durante, et altera g in nihilum plane redacta? Quae res profecto admiratione digna est! […]”; English translation – “Of the spots born the day before, only the sunspot “f” survived; the sunspot “g” had completely disappeared, until it did not leave even a faint trace. Therefore, the following question arises: why did this missing sunspot, as well as the other sunspot “f”, not remain on the Sun as long as the sunspot “c”, even though the previous day had the same size and an equally dense and compact matter? Why does the sunspot “c” remain and instead the sunspot “g” is reduced to nothing? It is a matter worth of wonder!”. Furthermore, Hevelius realized that some sunspots returned, *i.e.*, the same sunspot that disappeared at the solar limb appeared some days after on the opposite solar limb. As an example, we can cite the annotations made by Hevelius (1647, p. 521) on June 1644 about the great sunspot “d” represented in Figure 3A, which remained visible on the Sun for several solar rotations: original text – “[…] Eodemque simul tempore ad ortum, sub ipsa ferme Ecliptica, nova Macula e Faculis concomitata, denuo in obtutum venit; quam in numerum reducum referendam ac illam ipsam magnam egregiamque Maculam d, in imagine XX expressam esse evidentissime ex sequentibus colligo rationibus. Primo, quod Solem, altera vice, elapsis scilicet 27 diebus et quidem circa eandem fere peripheriae partem, denuo subingressa fuerit. Secundo, quod forma insuper plane eadem, puta, rotundata haloneque pariter circundata, sicut in priori cursu, in imagine videlicet XX apparuerit."; English translation – "[…] At that same time, towards the eastern limb almost under the same ecliptic, a new sunspot "e" accompanied by faculae, became visible again. I deduce that it is necessary to consider it within the returned sunspots, and that it is the same sunspot "d", big and imposing that appears in the image XX, and I reach such a conclusion for the following reasons. First, because it returned on the Sun, for the second time after 27 days and on the same part of the periphery. Second, because, in the image XX, it also had the same shape, that is, rounded and surrounded by a halo, as in its first path".

### 4.5. A greater contrast next to the solar limb

Finally, the last phenomenon observed by Hevelius that we want to show is the stronger contrast between the darkest and brightest regions near the solar limb. This is the "limb darkening" (Foukal, 2004), mentioned earlier by Galileo and Scheiner (2010). On 26 July 1643, Hevelius (1647, p. 506) commented: original text – "Denuo Faculae fulgentes circa Maculam e exortae sunt quae die subsequente multo ampliores luculentioresque factae sunt (quipped quod id plerumque prope horizontem fiery soleat) rariores vero et magis dispersae quo centro viciniores existunt, apparent"; English translation – "Again, there were bright faculae near the sunspot "e", which became much larger and brighter the next day (what, by the way, usually happens near the solar limb). Instead, they appear weaker and more scattered when they are closer to the center".

## 5. Estimation of solar activity

### 5.1. Solar activity level from the group number and active day fraction

The sunspot records published by Hevelius (1647) have been carefully analyzed in this work to obtain the number of single sunspots and sunspot groups recorded by Hevelius in each observation. We have defined the sunspot group according to the modern group classifications (McIntosh 1990). Thus, we have also calculated the monthly and annual number of active and quiet days. We consider one day as an active day when, at least, one sunspot was observed on the solar disk. On the contrary, when there is no sunspot on the Sun, we consider it as a quiet day. These data are publicly available on the website of the Historial Archive of Sunspot Observation (HASO, http://haso.unex.es). Table 1 presents the total monthly number of observation days (N), the monthly average of single sunspots (S) and sunspot groups (G), and the percentage of the active (AD) and quiet (QD) days with respect to the total monthly number of observations (N). We have corrected 41 of the 324 daily sunspot records (~13 %) in the Vaquero et al. (2016) data base that Hevelius (1647) made from 1642-1645. Among these 41 cases, we have corrected the group counting in 32 cases, wrong dates in 5 cases, and we have included 4 observations which were omitted in the group database. We note that we have considered 26 August 1644 as an active day. However, although one group was recorded by Hevelius according to the group database, we have not determined the number of groups (nor the individual sunspots) in that day because Hevelius recorded that he observed sunspots on the Sun that day without specifying numbers of groups or single sunspots and indicated that he could not perform the sunspot drawing. In Figure 6, we compare the monthly average of the group number obtained from the sunspot observations recorded by Hevelius (1647) calculated in this work (blue squares) and from the records included in the group database (red circles) for the same documentary source. We obtained a higher monthly average of the group number than the calculations made from the group database in 10 months and a lower monthly average only in October 1644. Thus the overall average daily group sunspot number for the revised Hevelius data base from 1642-1645 of 1.35 is 7 % higher than that (1.26) calculated from the Vaquero et al. (2016) data base. We note those averages were calculated from the daily group counting.

Table 1. Monthly number of days of observation (N), monthly average of the number of sunspots (S), and sunspot groups (G), and percentage of active (AD) obtained from Hevelius (1647).

| **YEAR** | **MONTH** | **N** | **S** | **G** | **AD (%)** |
|---|---|---|---|---|---|
| 1642 | 10 | 3 | 1.0 | 0.7 | 66.7 |
| 1642 | 11 | 14 | 2.0 | 1.1 | 92.9 |
| 1643 | 5 | 8 | 3.6 | 2.5 | 100.0 |
| 1643 | 6 | 11 | 8.0 | 2.9 | 100.0 |
| 1643 | 7 | 26 | 2.7 | 1.7 | 80.8 |
| 1643 | 8 | 30 | 3.2 | 1.4 | 76.7 |
| 1643 | 9 | 26 | 1.5 | 1.1 | 73.1 |
| 1643 | 10 | 20 | 1.8 | 0.8 | 70.0 |
| 1643 | 11 | 8 | 3.9 | 1.9 | 62.5 |
| 1643 | 12 | 6 | 1.2 | 1.2 | 100.0 |
| 1644 | 1 | 7 | 1.1 | 0.6 | 57.1 |
| 1644 | 2 | 7 | 5.9 | 2.7 | 100.0 |
| 1644 | 3 | 9 | 2.6 | 1.0 | 77.8 |
| 1644 | 4 | 26 | 3.9 | 1.8 | 69.2 |
| 1644 | 5 | 23 | 6.7 | 2.2 | 87.0 |
| 1644 | 6 | 13 | 2.2 | 1.5 | 92.3 |
| 1644 | 7 | 28 | 2.5 | 1.6 | 53.6 |
| 1644 | 8 | 16 | 0.5 | 0.4 | 37.5 |
| 1644 | 9 | 19 | 1.2 | 0.3 | 31.6 |
| 1644 | 10 | 6 | 1.2 | 1.0 | 66.7 |
| 1644 | 11 | 6 | 0.0 | 0.0 | 0.0 |
| 1644 | 12 | 8 | 0.0 | 0.0 | 0.0 |
| 1645 | 1 | 4 | 0.0 | 0.0 | 0.0 |

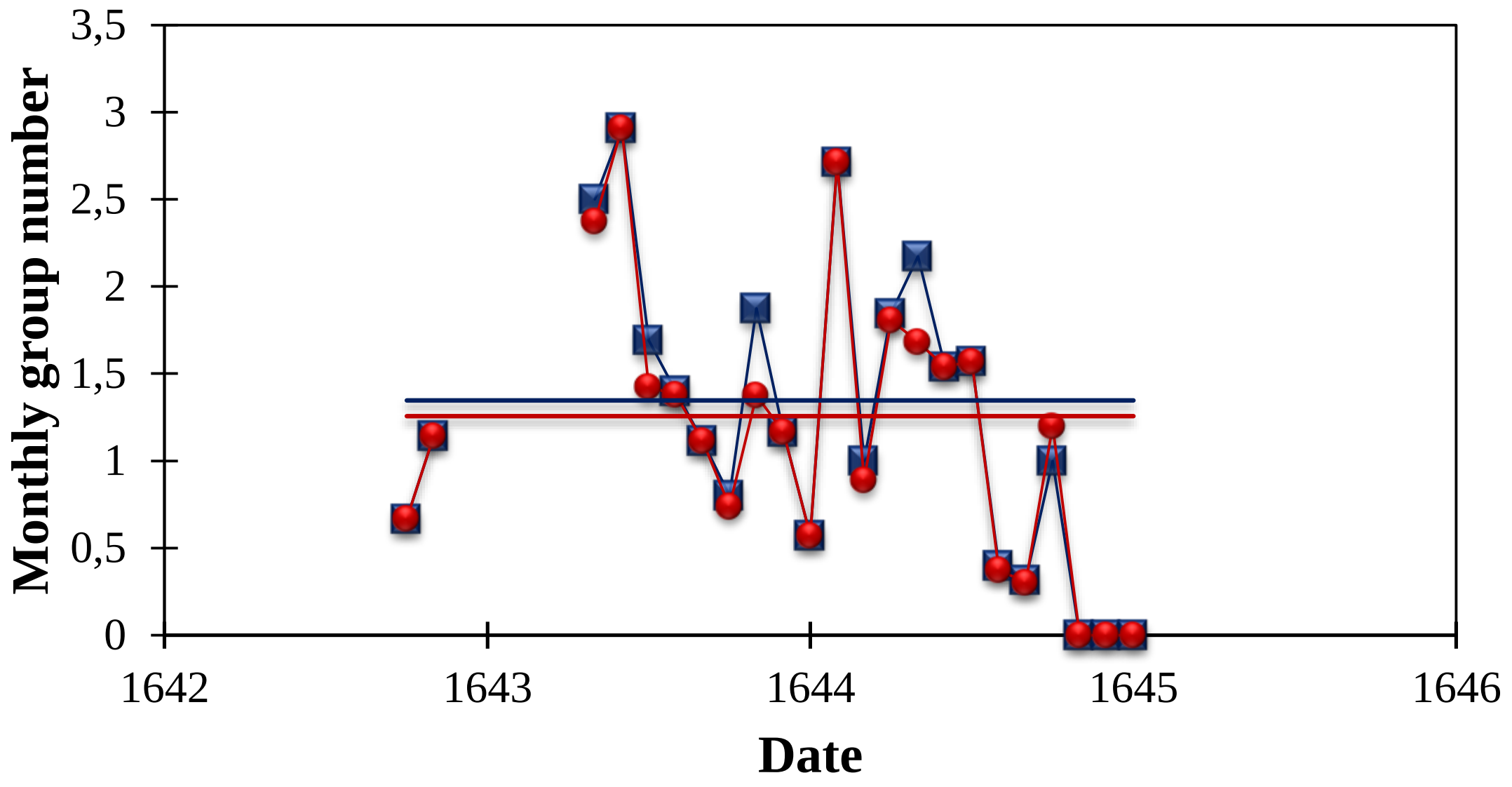

Figure 6. Monthly average of the group number according to this work (blue colors) and the group database (red colors) from the sunspot observations recorded by Hevelius

(1647). The horizontal lines represent the average of the group number for the whole period.

Recently, several studies have been carried out to evaluate the solar activity level during the MM from the number of active days (Kovaltsov et al. 2004; Carrasco & Vaquero 2016). The fraction of active days ($F_a$) is defined as the active day number divided by the total number of records. The value of the active day fraction calculated from the records included in Hevelius (1647), just prior to the MM, is significantly higher than the value obtained from observations made by Hevelius during the MM recorded in *Machina Coelestis* (Hevelius 1679). Figure 7 shows the active day fraction obtained in this work from the sunspot observations made by Hevelius during the period 1642-1645 and those one obtained by Carrasco et al. (2015) from the observations recorded by Hevelius during the period 1653-1675. The average of the active day fraction for the first observation period is approximately equal to 70 %, while it is about 30 % for the second period, during the first part of the MM. Moreover, Carrasco et al. (2015) calculated that the active day fraction during the Dalton Minimum, a period with a reduced solar activity that is not considered to be a grand minimum, is approximately 50 %. These facts suggest that solar activity was abnormally low during the MM and therefore this epoch can be really considered as a grand minimum of solar activity. However, we want to note some weakness of this conclusion. Our dataset does not have a good observational coverage during the MM. Furthermore, we are comparing sunspot observations carried out during four years (1642–1645) and exclusively in the declining phase of the solar cycle (Hevelius 1647) with data recorded during a 23-year period (Hevelius 1679) where we do not know exactly the phase of the solar cycles when those sunspot observations were recorded. We also emphasize that a maximum could have taken place in 1660 when a significant number of observations well distributed temporally throughout that year were recorded by Hevelius and the active day fraction obtained is less than 50 %.

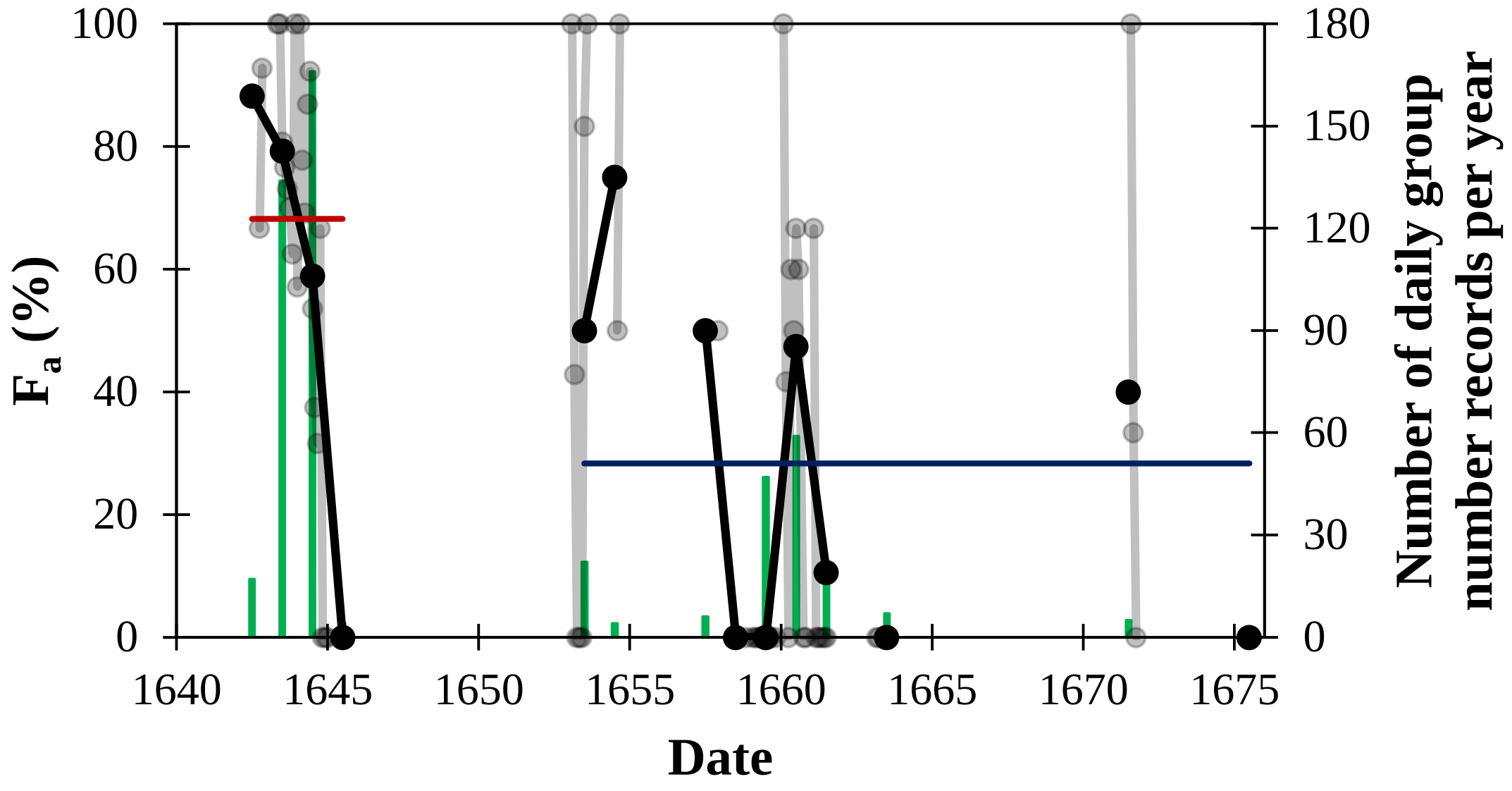

Figure 7. Active day fraction calculated according to sunspot observations made by Hevelius during the periods 1642-1645 (Hevelius 1647) and 1653-1675 (Hevelius 1679) is depicted by black (annual average) and grey (monthly average) dots. The average for the active day fraction during first (second) period is indicated by red (blue) horizontal line. Vertical green bars represent the number of annual observations.

5.2. Butterfly diagram

Recent efforts have been carried out in order to determine the sunspot positions recorded by historical observers. For example, Arlt (2009) and Arlt et al. (2013) calculated the sunspot positions according to the sunspot observations made by Staudacher in the 18th century and Schwabe during the 19th century, respectively. More recently, Karoff et al. (2019) have presented the construction of a record of sunspot positions made by Horrebow also in the 18th century. We have determined the heliographic coordinates for all sunspots recorded by Hevelius (1647), based on an ephemeris provided by the JPL Horizons system [footnote: https://ssd.jpl.nasa.gov/?horizons]. The analysis method is identical to the one used by Arlt et al. (2016). These data can be also consulted on the website: http://haso.unex.es. In addition, the heliographic coordinates have been cross-checked with the methodology proposed by Galaviz (2018) to obtain sunspot positions. Figure 8 depicts the butterfly diagram according to those calculations. All the sunspots range in latitudes between +20º and -20º, with 75%, approximately, of the sunspots located around 10º (between 5º and 15º N/S). It shows that this solar cycle (Solar Cycle -10) would be in its declining phase. Furthermore, the number of single sunspots recorded by Hevelius in the northern hemisphere (541) of the Sun is slightly greater than the number for the southern hemisphere (435). In this way, from the point of view of the probability of having this observed asymmetric distribution originated by random fluctuations (Vaquero et al. 2015b), the analysis shows that this probability is about 0.038 %. This p-value indicates that, at a significance level of 5 %, this asymmetry is statistically sound. We emphasize that this hemispheric behavior is different from that observed in the period 1670-1720 during the later phase of the MM when sunspots mainly appeared in the southern solar hemisphere (Ribes & Nesme-Ribes 1993).

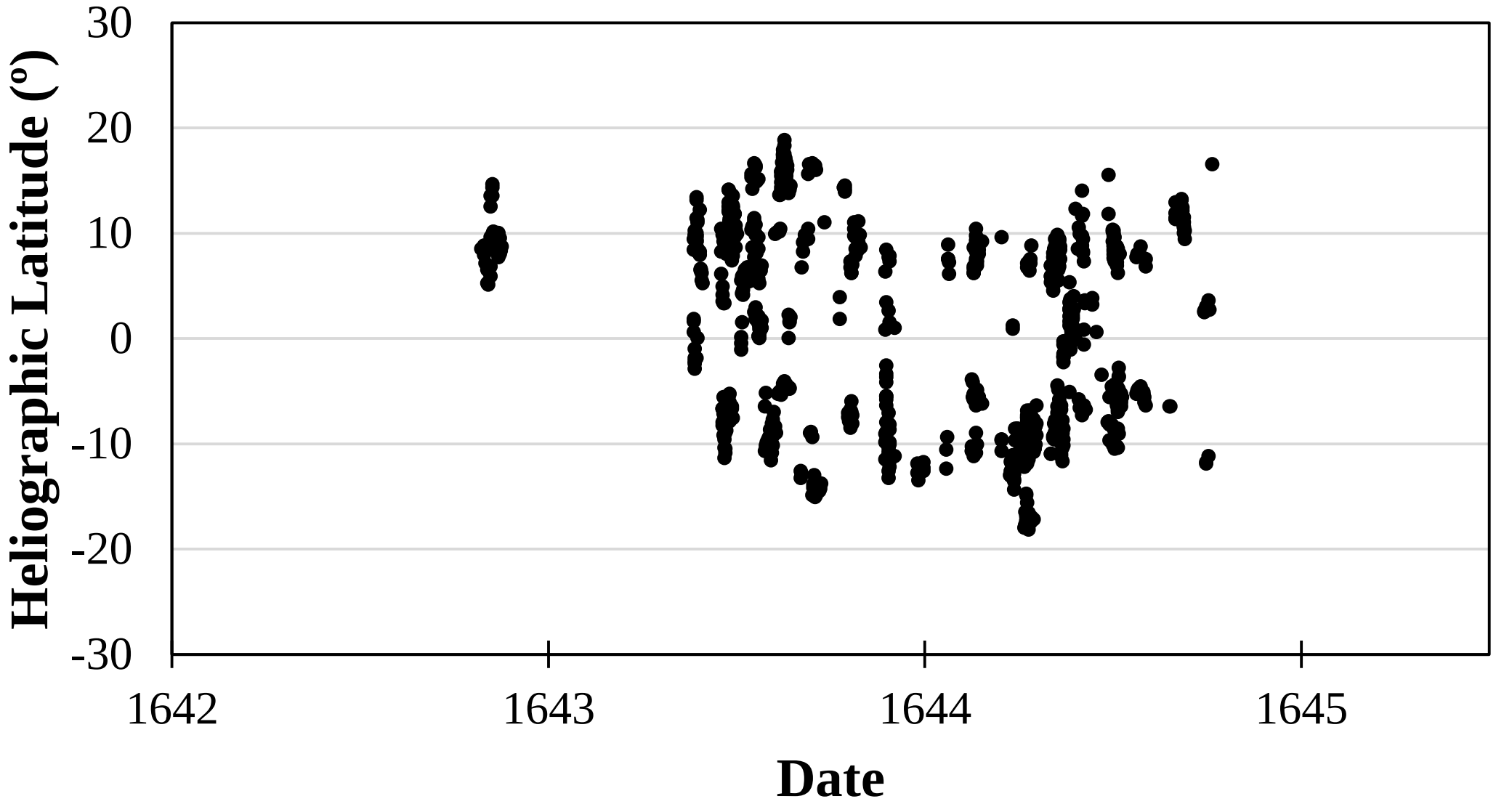

Figure 8. Butterfly diagram according to sunspot recorded by Hevelius (1647).

**6. Others significant comments**

In this section, we highlight three important comments made by Hevelius (1647, p. 95, 96, and 499) about his sunspot observations that show clearly the meticulousness of these records:

Original text – Page 95: "[…] 4. Faculae et maculae in Figuris repraesentatae, de die in diem, annuente coeli serenitate, sunt observatae; praeterea diligenter fuit determinata earum magnitudo, proportio, distantia, ut et color, densitas, nuclei, umbrae et cognatae res, ita ut omnia quae faciunt ad explicationem formae ac motus earum summo studio sint animadversa aerique incisa. […]"; English translation – Page 95: "[…] 4. Faculae and sunspots represented in the figures have been observed day to day, when allowed by a clear and calm sky. Furthermore, it has been carefully determined its magnitude, proportion and distance, as its color, density, nuclei, umbrae and other elements of this style, so it has been warned and noted with the greatest care everything what is related with the description of its aspect and movement […]".

Original text – Pages 95-96: "[…] I. An Maculae sint dicendae stellae, quae circa Solem suum motum exerceant eumque constanter ac semper observant? Nam, quid ad hanc quaestionen respondendum sit, ex observationibus haud obscure patet. Quod enim nequeant esse stellae, ex his colligitur. Primum, Maculae nunquam fere apparent rotundae, stellae vero simper. 2. Deinde si essent stellae, non tam miras subirent mutationes, nec iam maiores, iam minores conspicerentur, sed eandem formam obtinerent. 3. Oporteret illas singulis diebus et omni tempore reverti et nunquam evanescere. At hoc non fit. Quare recte infertur maculas non esse stellas. […]"; English translation – Pages 95-96: "[…] Can sunspots be considered stars that move around the Sun and that they constantly and always observe it? The answer to this question is clearly evident from the observations, so from them it follows that they cannot be stars. First, the spots almost never are shown as a circle, instead the stars always. 2. Secondly, if they were stars, they would not experience such amazing changes, or they would sometimes not appear larger, sometimes smaller, but they would have the same shape. 3. It would be imperative that they returned every day and all the time, and that they will never dissipate, and this fact does not happen, from which it can be deduced that the sunspots are not stars […]". Therefore, we can conclude that Hevelius did not consider that sunspots were stars (or celestial bodies) orbiting the Sun unlike other astronomers in that time (Carrasco et al., 2019b).

Original text – Page 499: "[…] in medium proferamus; ut ita quilibet, isque Scientiae sidereae cupidus, ex oculari talium observationum inspectione ac demostratione, omnia et singula, quaecunque in praedicto capit de macularum ortu et occasu, cum Astronomico tum Physico, illarumque diversis generibus, et ut vario, constantia tamen mirabili motu aliisque huc pertinentibus rebus, prolata fuerint, eo melius faciliusque percipiat; tandem aliquando nobiscum statuens (haud obstante Peripateticorum inveterato dogmate) generationem ac corruptionem, augmentationem ac diminutionem, et quidem nonnunquam sane admirabiliter magnam, in ipsis coeli visceribus existere, quemadmodum id ex observatis hactenus observandisque nunquam non periodis Macularum Facularumque Solarium, Solis luce clarius elucebit. […]"; English translation – Page 499: "[…] Thus, anyone, mainly those passionate about astronomy, understands better from the ocular inspection and description of such observations each one of the questions that were mentioned in the chapter cited about the appearance and concealment of sunspots, both astronomic and physical, and on the different types of sunspots and their movement by an admirable continuity, although varied, and on other issues related with the topic; and finally, sometime (without the ancient dogma of the pathetic [ridiculous] people being an obstacle) establishing with us that its creation and alteration, increase and decrease, admirably big sometimes, exist in the very depths of the sky, in the same way as from the periods of spots and faculae observed hitherto and that will always be observed, will shine more clearly than the light of the Sun […]".

Some recent works (Rek 2013; Zolotova & Ponyavin 2015) claimed that the MM was not a grand minimum of solar activity but a secular minimum. For example, Zolotova & Ponyavin (2015) pointed out that during the 17th century most of the sunspots recorded in the drawings tend to be circularized as consequence of the ideas of that time as the Aristotelian geocentrism and a perfect and immaculate Sun. Moreover, Zolotova & Ponyavin (2015) proposed that objects on the solar surface with an irregular shape or small sunspots could have been omitted in a textual report because it was impossible to recognize that these objects were celestial bodies. As an example, the sunspot drawings made by Hevelius (1647) were used to indicate that Hevelius was probably familiar with the works of Scheiner (Scheiner 1630) and his sunspot drawings published in *Selenographia* tend to be similar to a set of circles. However, according to Hevelius' annotations showed in *Selenographia*, Hevelius endeavored to draw and describe every detail of the sunspots observed by him without taking into account the ancient dogmas about the Sun.

As a final comment, we want to emphasize that Hevelius was the first observer to realize and record the absence of sunspots at the onset of MM:

Original text – "Ad diem usque 5 Nov. ob aerem nimis turbulentum nihil annotatum neque animadversum. Sic et die 14, 25, 28, 29, 30, 2 Decemb., 3, 4, 5, 6, 11, 16, 21, ut et 5 Ianuar. An. 1645, 6, 13 et 24, Sol plane ab omnibus Maculis defaecatus extitit. Estque sane res admirabilis quod in tam longo temporis intervallo nequicquam prodierit. Utrum vero in diebus intermediis, in quibus ob tristem Coeli vultum nobis observare minime obtigit, aliquid in Sole emerserit nec ne, profecto me clam est. Interim tamen id scire magnopere interest. Quamobrem contemplators phaenomenorum coelestium quoscunque etiam atque etiam rogatos velim, siquidem tum temporis vigiles in Solem direxerint oculos, ut nos de his certiores reddant. Etenim, si illis item supra dictis intermediis diebus, quicquam oculis animadvertere negatum, tunc certe absque haesitatione pronunciare licebit Solem per tres integros menses mundum purumque ab omni Macula Faculaque esse spectatum. Id quod observari maxime omnino meretur. Eo praesertim cum id ipsum, quantum sciam, nunquam adhuc a quopiam fuerit deprehensum. [...]"; English translation – "Until 5 November, as the atmosphere was too turbulent, no observations can be recorded. Thus, the Sun totally appeared clean of sunspots on 14, 25, 28, 29 and 30 November 1644, 2, 3, 4, 5, 6, 11, 16 and 21 December 1644, and 5, 6, 13 and 24 January 1645. And it is very admirable in such a long interval of time that nothing was appreciated. Whether in the intermediate days, in which we did not have any opportunity to observe because of the dark aspect of the sky, anyone observed something on the Sun or not, I certainly do not know it. However, it is extremely important to make it known. Thus, I would like that this question was confirmed by all those observers of the celestial phenomena that I have asked over and over again, if they, by chance, were attentive at that time and fixed their eyes on the Sun. Because if they have not been able to observe anything with their own eyes in regard to the above-mentioned intermediate days, then one can clearly and undoubtedly say that the Sun was clear of sunspots for three whole months. And this fact deserves to be considered very highly, first of all, because nobody, as far as I know, has noticed it […]".

As it can be seen in this last comment made by Hevelius, he recorded no sunspots during three whole months, being the first astronomer to realize of the absence sunspots during the 17th century. Nevertheless, Hoyt & Schatten (1998) indicated that a sunspot was observed by Linemanns, an observer unknown by Wolf, during the 3-month period in which Hevelius did not recorded sunspots. In particular, according to Hoyt & Schatten (1998), Linemanns observed one sunspot on 15 November 1644, one day after an observation with zero sunspots record by Hevelius. We have consulted the letters sent by Linemanns to Hevelius at that time and after translating from Latin, for that date, we have only found records about an occultation of Aldebaran by the Moon. It is true that Linemanns pointed out a "macula D" in the text but it is the *Mare Crisium* of the Moon. While, we cannot guarantee that the sunspot observation made by Linemanns included in the Hoyt & Schatten (1998) database is wrong because Hoyt and Schatten did not provide the specific reference for this record, after checking the correspondence of Linemanns, we question the validity of that sunspot record made by Linemanns on 15 November 1644. In fact, we think that it could be a mistake in the data compilation by Hoyt & Schatten (1998). Therefore, taking into account the sunspot records of Hevelius, it seems that this 3-month period occurred at the end of 1644 and in the beginning of 1645 could be considered the onset of the MM.

Furthermore, it is worth mentioning that, except for the MM, 3-month intervals without sunspots on the Sun have only occurred (in the last four centuries) in 1755, and in the Dalton Minimum (1798, 1809–1812) (Vaquero et al. 2016). We note that the observational coverage in those periods was low (57.0 % in 1755, 23.0 % in 1798 and 34.4 % for the period 1809–1812) and there are long periods of consecutive days without observations during those 3-month periods (one period of 18 consecutive days without observation in 1755, one case of 22 and other of 49 in 1798, multiple cases greater than 10 and 20 consecutive days without observations in 1809, 1811, and 1812 and only in 1810 the maximum number of consecutive days without observations is less than 10). Instead, considering sunspot area measurements from April 1874 to January 2017 made at the Royal Greenwich Observatory and Debrecen Observatory, if we discard those sunspots with areas lower than 10 millionths of a solar hemisphere, a 3-month period without sunspots can be found in the minimum of the Solar Cycle 15 (in 1913) where the observational coverage is equal to 100 %. Furthermore, if we do not consider sunspots with areas lower than 20 millionths of a solar hemisphere, a 3-month period without sunspots can be also located in the minima of the solar cycles 14 (in 1901, almost 100 % of observational coverage) and 23 (in 2009, around 90 % of observational coverage). Although the number of observation made by Hevelius during that period of three months is not very numerous (18 days in total), it should be noted that these observations have a good temporal distribution. The longest data gaps were 10, 11, and 14 days, i.e. only the two observations between 21 December and 5 January have a difference greater than 13 days, one half of the synodic rotation period of the Sun (Heristchi & Mouradian 2009). Because this 3-month period coincided with the end of the previous cycle, we cannot rule out that small, short-lived sunspots, which are common near solar minimum, were still present on the Sun but we can discard that long-lived sunspots were present on the Sun at the onset of the MM. We also note that Hevelius (1679) recorded observations with zero sunspots during eight consecutive months in the MM, from February to September 1659 (Carrasco et al. 2015). The average between two consecutive days for those observations is equal to 4.8 days and, among the 45 total records made by Hevelius in that period, the difference between two consecutive records was larger than 13 days (approximately half of the synodic rotation period of the Sun) in only two cases. We acknowledge that the length of this spotless period might have been affected by the poor coverage (approximately 20%) caused by bad weather in Europe during the MM years and that short-lived sunspots could have been missed.

## 7. Conclusions

Johannes Hevelius recorded in *Selenographia* his sunspot observations made at Danzig (currently Gdansk, Poland) during the period 1642–1645. He presented his records in sunspot drawings, along with annotations for each observing day. The importance of these observations is that they are the only systematic sunspot observations just before the MM. Although this documentary source was previously studied by Hoyt & Schatten (1998), we present a new analysis after translating all the relevant information about sunspots from the original Latin texts.

Hevelius improved the helioscope, an instrument invented by Christoph Scheiner to observe the Sun, in order to allow a single person to carry out solar observations alone. We have not found specific information about the focal length and aperture of the helioscope used by Hevelius to observe sunspots. However, we can speculate that the diameter of the lens would be around 0.05 m since that was the diameter of the brass tube where telescopes were placed. Moreover, Kampa (2018) indicated the focal length was about 0.6 m (2 feet). We highlight that the quality of the methodology used by Hevelius was appreciated by other astronomers as, such as, Gassendi and Linemann.

Hevelius carried out meticulous and exceptional sunspot observations during that time. He recorded sunspots and faculae and other less common phenomena defined by him as “clouds”, “fogs”, “umbrae”, and “halos”. Halos are the most striking phenomenon recorded by Hevelius. We have not identified a modern counterpart for this phenomenon and it is unclear how Hevelius was able to observe them with his helioscope. The remaining phenomenon could refer to faculae or the darker regions between sunspots due to a contrast effect between the faculae and photosphere. Moreover, Hevelius analyzed the formation and evolution of sunspots and realized the path of the sunspots crossing the solar disc depended on the month of the year and some sunspots returned for several solar rotations. He also described a stronger contrast between the darkest and brightest regions next to the solar limb.

We present a new recount of the daily sunspot groups recorded by Hevelius, in addition to the daily number of individual sunspots. We found that the solar activity level calculated from the group number is 7 % higher than that one obtained from the group database considering the same records. Moreover, the active day fraction calculated in this work from the sunspot records made by Hevelius (1647) just before the MM (70 %) is significantly higher than the active day fraction obtained from the sunspot observation made by Hevelius (1679) during the MM (30 %). In addition, Carrasco et al. (2015) calculated that the active day fraction during the Dalton Minimum (50 %) is also significantly higher than the level of solar activity that they obtained considering the Hevelius’ records made during the MM. This suggests that the solar activity level during the MM is compatible with a grand minimum period. However, we are aware that our analysis presents some limitations because our dataset does not have a good observational coverage. We have also calculated the positions for all sunspots recorded by Hevelius (1647) showing the butterfly diagram. All the sunspots range in latitudes between +20º and -20º, making it likely that the solar cycle was in its declining phase. Furthermore, a slightly larger number of sunspots was observed in the northern hemisphere of the Sun (541 *versus* 435). Thus, the strong hemispheric asymmetry recorded during the MM did not exist just before that period. Therefore, other parameters should be analyzed to find precursors of a grand minimum period.

Finally, from the annotations made by Hevelius, we want to highlight that he carefully recorded everything that he observed and he did not think that sunspots were celestial bodies orbiting the Sun. Therefore, the argument hypothesized by Zolotova & Ponyavin (2015) that the sunspot drawings published by Hevelius in *Selenographia* tend to be a set of circles is erroneous. On the other hand, Hevelius recorded only spotless days from November 1644 to January 1645 and was the first astronomer who noticed the absence of sunspots on the Sun during an extended period although we note that Marius (1619) wrote: "… that now I have seen much fewer during the last 1.5 years, often not a single one, something that hasn't hapened in the years before". We note that there were only 18 observations during these 3 months because of poor weather, they were well separated in time and we can discard the possibility that long-lived sunspots appeared on the Sun in that time. Thus, the onset of the MM could be set in that 3-month period without reported sunspots. Hevelius was a meticulous observer who provided one of the best sunspot observation set made in the first years of the telescopic era.

**Acknowledgements**

This work was partly funded by FEDER-Junta de Extremadura (Research Group Grant GR18081, GR18097 and project IB16127) and from the Ministerio de Economía y Competitividad of the Spanish Government (CGL2017-87917-P). VSP acknowledges the support from Deutsche Forschungsgemeinschaft by grant no. AR 355/12-1. Authors have benefited from the participation in the ISSI workshops led by Owens and Clette on calibration of the sunspot number. Authors acknowledge the helpful discussion with Héctor Socas Navarro about Section 4.

**Disclosure of Potential Conflicts of Interest**

The authors declare that they have no conflicts of interest.

**References**

Abarbanell, C., Wöhl, H. 1981, Solar Rotation Velocity as Determined from Sunspot Drawings of Hevelius, J. in the 17TH-CENTURY, Solar Phys., 70, 197. doi: 10.1007/BF00154400.

Arlt, R. 2009, The Butterfly Diagram in the Eighteenth Century, Solar Phys., 255, 143. doi: 10.1007/s11207-008-9306-5.

Arlt, R., Senthamizh Pavai, V., Schmiel, C., Spada, F. 2016, Sunspot positions, areas, and group tilt angles for 1611-1631 from observations by Christoph Scheiner, A&A, 595, A104. doi: 10.1051/0004-6361/201629000.

Arlt, R., Leussu, R., Giese, N., Mursula, K., Usoskin, I. G. 2013, Sunspot positions and sizes for 1825-1867 from the observations by Samuel Heinrich Schwabe, MNRAS, 433, 3165. doi: 10.1093/mnras/stt961.

Carrasco, V. M. S., Vaquero, J. M. 2016, Sunspot Observations During the Maunder Minimum from the Correspondence of John Flamsteed, Solar Phys., 291, 2493. doi: 10.1007/s11207-015-0839-0.

Carrasco, V. M. S., Villalba Álvarez, J., Vaquero, J. M. 2015, Sunspots During the Maunder Minimum from *Machina Coelestis* by Hevelius, Solar Phys., 290, 2719. doi: 10.1007/s11207-015-0767-z.

Carrasco, V. M. S., Vaquero, J. M., Gallego, M. C. 2018, Could a Hexagonal Sunspot Have Been Observed During the Maunder Minimum? Solar Phys., 293, 51. doi: 10.1007/s11207-018-1270-0.

Carrasco, V. M. S., Vaquero, J.M., Gallego, M.C., Villalba Álvarez, J. Hayakawa, H. 2019a, Two debatable cases for the reconstruction of the solar activity around the Maunder Minimum: Malapert and Derham, MNRAS, 485, L53. doi: 10.1093/mnrasl/slz027.

Carrasco, V. M. S., Gallego, M.C., Villalba Álvarez, J., Vaquero, J.M. 2019b, Sunspot observations by Charles Malapert during the period 1618–1626: a key data set to understand solar activity before the Maunder minimum, MNRAS, 488, 3884. doi: 10.1093/mnras/stz1867.

Clette, F., Svalgaard, L., Vaquero, J.M., Cliver, E.W. 2014, Revisiting the Sunspot Number, Space Sci. Rev., 186, 35. doi: 10.1007/s11214-014-0074-2.

Eddy, J. A. 1976, The Maunder Minimum, Science, 192, 1189. doi: 10.1126/science.192.4245.1189.

Foukal, P. V. 2004, Solar Astrophysics (2nd ed.; New York, Wiley).

Galaviz, P. 2018, Determinación y análisis de regiones activas solares en los últimos siglos, (Mérida, University of Extremadura).

Galilei G., Scheiner C. 2010, On Sunspots (Chicago: The University of Chicago Press).

Gómez, J. M., Vaquero, J. M. 2015, The sunspot observations by Rheita in 1642, The Observatory, 135, 220.

Heristchi, D., Mouradian, Z. 2009, The global rotation of solar activity structures, A&A, 497, 835. doi: 10.1051/0004-6361/200809582.

Hevelius, J. 1647, Selenographia: sive lunae description (Danzig: Hünefeld).

Hevelius, J. 1679, Machina Coelestis Pars Posterior (Danzig: Simon Reiniger).

Hoyt, D. V., Schatten, K. H. 1998, Group Sunspot Numbers: A New Solar Activity Reconstruction, Solar Phys., 179, 189. doi: 10.1023/A:1005007527816.

Kampa, I. 2018, Die astronomischen Instrumente von Johannes Hevelius (Hamburg: Tredition).

Karoff, C., Jørgensen, C. S., Senthamizh Pavai, V., Arlt, R. 2019, Christian Horrebow's Sunspot Observations - II. Construction of a Record of Sunspot Positions, Solar Phys., 294, 78. doi: 10.1007/s11207-019-1466-y.

Kovaltsov, G. A., Usoskin, I. G., Mursula, K. 2004, An Upper Limit on Sunspot Activity During the Maunder Minimum, Solar Phys., 224, 95. doi: 10.1007/s11207-005-4281-6.

Linemann, M. A. 1654, Delicae Calendario-Graphicae, das ist die sinnreichsten und allerkünstlichsten Fragen und Antwork: Darinnen die edelsten Geheimnüsse der Physic, Astronomi, AStrologi, Geographi, etc. Etc. (Königsberg: Mense).

Marius, S. 1619, Astronomische und Astrologische Beschreibung des Kometen von 1618 (Nürnberg: Johann Lauer).

Maunder, E. W. 1890, A Prolonged Sunspot Minimum, MNRAS, 50, 251.

McIntosh, P. S. 1990, The classification of sunspot groups, Solar Phys., 125, 251. doi: 10.1007/BF00158405.

Mueller, P. R. 2000, An Unblemished Success: Galileo's Sunspot Arguement in the Dialogue, Journal for the History of Astronomy, 31, 279.

Muñoz-Jaramillo A., Vaquero J. M., 2018, Nat. Astron., 3, 205, DOI: 10.1038/s41550-018-0638-2.

Olhoff, J. E. 1683, Excerpta ex Literis Illustrum, et Clarissimorum Virorum Ad Nobilissimum, Ampliss. Et Consultiss. DN. Johannem Hevelium Cons. Gedanensem (Gdansk: Janssonius Waesbergius).

Rast, M. P., Meisner, R. W., Lites, B. W., Fox, P. A., White, O. R. 2001, Sunspot Bright Rings: Evidence from Case Studies, ApJ, 557, 864. doi: 10.1086/321673.

Rast, M. P., Fox, P.A., Lin, H., Lites, B.W., Meisner, R.W., White, O.R. 1999, Bright rings around sunspots, Nature, 401, 678. doi: 10.1038/44343.

Rek, R. 2013, in Johannes Hevelius and His World: Astronomer, Cartographer, Philosopher and Correspondent, ed. R. L. Kremer & J. Wlodarczyk (Warsaw: Polish Academy of Sciences), 81.

Ribes, J. C., Nesme-Ribes, E. 1993, The solar sunspot cycle in the Maunder minimum AD1645 to AD1715, A&A, 276, 549.

Scheiner, C. 1630, Rosa Ursina Sive Sol (Brassiano: Andrea Fei).

Smith, A. M. 1985, Galileo's proof for the Earth's motion from the movement of sunspots, Isis, 76, 543.

Spoerer, G. 1889, Über die Periodicität der Sonnenflecken seit dem Jahre 1618, vornehmlich in Bezug auf die heliographische Breite derselben, und Nachweis einer erhebliche Störung dieser Periodicität während eines langen Zeitraumes (Halle: Blochmann).

Svalgaard, L., Schatten, K. H. 2016, Reconstruction of the Sunspot Group Number: The Backbone Method, Solar Phys., 291, 2653 doi: 10.1007/s11207-015-0815-8.

Usoskin, I. G. 2017, A history of solar activity over millennia, Living Rev. Solar Phys., 14, 3. doi: 10.1007/s41116-017-0006-9.

Usoskin, I. G., Arlt, R., Asvestari, E., et al. 2015, The Maunder minimum (1645–1715) was indeed a Grand minimum: A reassessment of multiple datasets, A&A., 581, A95. doi: 10.1051/0004-6361/201526652.

Vaquero, J. M., Vázquez, M. 2009, The Sun Recorded Through History (Berlin: Springer).

Vaquero, J. M., Gallego, M. C. 2014, Adv. Space Res., 53, 1162. doi: 10.1016/j.asr.2014.01.015.

Vaquero, J. M., Nogales, J. M., Sánchez-Bajo, F. 2015b, Sunspot latitudes during the Maunder Minimum: A machine-readable catalogue from previous studies, Adv. Space Res., 55, 1546. doi: 10.1016/j.asr.2015.01.006.

Vaquero, José M., Gallego, M. C., Usoskin, Ilya G., Kovaltsov, Gennady A. 2011, Revisited Sunspot Data: A New Scenario for the Onset of the Maunder Minimum, ApJ Lett., 731, L24. doi: 10.1088/2041-8205/731/2/L24.

Vaquero, J. M., Kovaltsov, G. A., Usoskin, I. G., Carrasco, V. M. S., Gallego, M. C. 2015a, Level and length of cyclic solar activity during the Maunder minimum as deduced from the active-day statistics, A&A, 577, A71. doi: 10.1051/0004-6361/201525962.

Vaquero, J. M., Svalgaard, L., Carrasco, V. M. S., Clette, F., Lefèvre, L., Gallego, M. C., Arlt, R., Aparicio, A. J. P., Richard, J. -G., Howe, R. 2016, A Revised Collection of Sunspot Group Numbers, Solar Phys., 291, 3061. doi: 10.1007/s11207-016-0982-2.

Waldmeier, M. 1939, Über die Struktur der Sonnenflecken, MiZur, 14, 439.

Zolotova, N. V., Ponyavin, D. I. 2015, The Maunder Minimum is Not as Grand as it Seemed to be, ApJ, 800, 42. doi: 10.1088/0004-637X/800/1/42.